\documentclass[10pt]{emulateapj}
\usepackage{apjfonts}
\usepackage{epsfig}
\def\Meszaros{M\'esz\'aros~}

\begin{document}

\title{Klein-Nishina effects on the high-energy afterglow emission of gamma-ray bursts }

\author{Xiang-Yu  Wang\altaffilmark{1},  Hao-Ning He\altaffilmark{1}, Zhuo Li\altaffilmark{2,3}, Xue-Feng Wu\altaffilmark{4,5} and Zi-Gao Dai\altaffilmark{1} }
\altaffiltext{1}{Department of Astronomy, Nanjing University,
Nanjing 210093, China}\altaffiltext{2}{Department of Astronomy,
Peking University, Beijing 100871, China}\altaffiltext{3} {Kavli
Institute for Astronomy and Astrophysics, Peking University,
Beijing 100871, China} \altaffiltext{4}{Department of Astronomy
and Astrophysics, Pennsylvania State University, 525 Davey Lab,
University Park, PA 16802}\altaffiltext{5}{Purple Mountain
Observatory, Chinese Academy of Sciences, Nanjing 210008, China}

\begin{abstract}
Extended high-energy($\ga100$ MeV) gamma-ray emission that lasts
much longer than the prompt sub-MeV emission has been detected
from quite a few gamma-ray bursts (GRBs) by Fermi Large Area
Telescope (LAT) recently. A plausible scenario is that this
emission is the afterglow synchrotron emission produced by
electrons accelerated in the forward shocks. In this scenario, the
electrons that produce synchrotron high-energy emission also
undergo inverse-Compton (IC) loss and the IC scattering with the
synchrotron photons should be in the Klein-Nishina regime. Here we
study effects of the Klein-Nishina scattering on the high-energy
synchrotron afterglow emission. We find that, at early times the
Klein-Nishina suppression effect on those electrons that produce
the high-energy emission is usually strong and therefore their
inverse-Compton loss is small with a Compton parameter $Y\la {\rm
a ~few}$  for a wide range of parameter space. This leads to a
relatively bright  synchrotron afterglow at high energies that can
be detected by Fermi LAT. As the Klein-Nishina suppression effect
weakens with time, the inverse-Compton loss increases and could
dominate over the synchrotron loss in some parameter space. This
will lead to a faster temporal decay of the high-energy
synchrotron emission than what is predicted by the standard
synchrotron model, which may explain the observed rapid decay of
the early high-energy gamma-ray emission in GRB090510 and
GRB090902B.

\end{abstract}


\section{Introduction}
With the launch of Fermi satellite, one of the new features of
gamma-ray bursts at high-energies  has been established, i.e. GRBs
show extended high-energy ($\ga100$ MeV) emission which lasts much
longer than the prompt phase. This extended emission has been seen
in both long and short GRBs and the flux usually decays with time
after the initial peak. In some cases (e.g. GRB090510 and
GRB090902B), the temporal decay is a simple power-law decay with a
slope ranging from $-1.3$ to $-1.5$ (Abdo et al. 2009a,b;
Ghirlanda, Ghisellini, Nava 2009; De Pasquale et al. 2009). One of
the proposed models for such emission is the hadronic cascade
emission model, in which the high-energy photons produced by the
accelerated ultra-high energy protons can not escape the soft
photon field and a cascade is induced (Abdo et al. 2009b). This
model has been applied to the extended emission of GRB941017
(Dermer \& Atoyan 2004), but whether it can explain the simple
power-law decay of the Fermi LAT bursts is unknown. The long-lived
behavior and not very rapid decay of the high-energy emission from
GRB090510 and GRB090902B can not be easily explained by  the
reverse shock emission model either (Wang et al.
2001a,b){\footnote{The decay of the high-energy emission in
GRB080825C is  steeper than $t^{-1.7}$ (Abdo et al. 2009c), which
is consistent with the synchrotron self-inverse Compton scattering
emission from the reverse shock or cross IC scatterings between
the reverse shock and forward shock (Wang et al. 2001a,b; Granot
\& Guetta 2003; Pe'er \& Waxman 2004; Wang et al. 2005).}}. On the
other hand, the simple power-law decay with a modestly large slope
is reminiscent of the afterglow emission. The self inverse-Compton
(IC) emission of the afterglow has been long thought to produce a
high-energy component (e.g. Zhang \& \Meszaros 2001; Sari \& Esin
2001; Fan et al. 2008; Gou \& \Meszaros 2007 ), but the light
curve is expected to rise initially and start to decay minutes to
hours after the burst. Kumar \& Barniol Duran (2009a) proposed
that the extended high-energy emission from GRB080916C is due to
afterglow synchrotron emission. This mechanism has also been
proposed to explain the extended high-energy emission from
GRB090510 (Gao et al. 2009; Ghirlanda, Ghisellini, Nava 2009;
Ghisellini, Ghirlanda, Nava 2009, De Pasquale et al. 2009) and
GRB090902B (Kumar \& Barniol Duran 2009b).

In the latter synchrotron afterglow scenario, the high-energy
extended emission is produced by the electrons  in the forward
shock via synchrotron emission. The shock-accelerated electrons
are usually assumed to have a power-law form in energy
distribution, i.e. $dN_e/d\gamma_e\propto \gamma_e^{-p}$, where
$\gamma_e$ is the Lorentz factor of electrons. These electrons
also suffer IC loss by scattering synchrotron photons. Due to that
the scattering between large $\gamma_e$ electrons and the
synchrotron photons could enter the Klein-Nishina (KN) scattering
regime, higher energy electrons may suffer smaller IC loss and as
a consequence, their synchrotron emission is stronger. Since the
Lorentz factor of the electrons producing the high-energy
afterglow emission are usually large, the KN scattering effect
must be taken into account when one calculate the synchrotron
high-energy afterglow emission. In this paper, we study the effect
of the KN scattering on the high-energy afterglow emission of
GRBs. Recently, Wang et al.(2009) studied the KN effect on the
prompt emission spectrum of GRBs and Nakar et al. (2009) studied
the KN effect on optically thin synchrotron and synchrotron
self-Compton spectrum in general. In this paper, we focus on the
early high-energy afterglow emission and confront the theoretical
results  with the high-energy afterglow observations by Fermi LAT.

The paper is organized as follows. First we study how the KN
scattering affects the electron distribution in the forward shock
in $\S$ 2. Then in $\S$ 3 we calculate the Compton parameters for
the high-energy electrons that produce high-energy gamma-ray
emission and study their evolution with time. With the Compton
parameters known, we calculate the light curves of high-energy
afterglows in $\S$ 3. Finally, we summarize  our findings in In
$\S$ 4.
\section{Klein-Nishina effect on the electron distribution}
We define the Compton parameter for electrons with Lorentz factor
$\gamma_e$ as the ratio of the synchrotron self-inverse Compton
(SSC) to the synchrotron  emissivity, i.e.
\begin{equation}
Y(\gamma_e)=\frac{P_{ssc}(\gamma_e)}{P_{syn}(\gamma_e)}.
\end{equation}
When KN effects  are unimportant, i.e. the IC scattering of
$\gamma_e$ electrons with synchrotron photons are in the Thomson
scattering regime,  $Y(\gamma_e)$ is a constant and its value has
been derived by Sari \& Esin (2001) for the GRB afterglow.
However, for high energy electrons whose KN effect become
important, $Y(\gamma_e)$ depends on $\gamma_e$. We approximate
\begin{equation}
Y(\gamma_e)=\frac{U_{\rm syn}[\nu<\nu_{KN}(\gamma_e)]}{U_B}
\end{equation}
for the afterglow emission (see also Li \& Waxman 2006), where
$h\nu_{KN}(\gamma_e)=\Gamma m_e c^2/\gamma_e$ is the critic energy
of scattering photons above which the scatterings with electrons
of energy $\gamma_e$ just enter the KN scattering regime ($\Gamma$
is the bulk Lorentz factor of the emission region), $U_{\rm
syn}[\nu<\nu_{KN}(\gamma_e)]$ is the energy density of the
synchrotron photons with frequency below $\nu_{KN}(\gamma_e)$ and
$U_B$ is the energy density of the magnetic field.

The KN effect can affect the electron distribution, which is given
by
\begin{equation}
N(\gamma_e)=\left\{{}
\begin{array}{ll}
C_1\gamma_e^{-p} \,\,\,\,\,\,\,\,\,\,\,\,\,\,\,\,\,\,\,\,\,\,\,\,\,\,\,\,\,\,\,\,\,\,\,\, \gamma_m<\gamma_e<\gamma_c \\
\frac{1+Y(\gamma_c)}{1+Y(\gamma_e)}C_1\gamma_c \gamma_e^{-p-1}
\,\,\,\,\,\, \gamma_c<\gamma_e
\end{array} \right .
\end{equation}
for the slow-cooling  case and
\begin{equation}
N(\gamma_e)= \frac{C_2}{1+Y(\gamma_e)}\left\{{}
\begin{array}{ll}
\gamma_e^{-2} \,\,\,\,\,\,\,\,\,\,\,\,\,\,\,\,\,\,\,\,\,\,\,\,\,\,\,\ \gamma_c<\gamma_e<\gamma_m\\
\gamma_m^{p-1}\gamma_e^{-p-1}
\,\,\,\,\,\,\,\,\,\,\,\,\,\,\,\,\,\,\,\,\,\,\,\,\
\gamma_m<\gamma_e
\end{array} \right .
\end{equation}
for the fast-cooling  case respectively (Nakar et al. 2009), where
$C_1$ and $C_2$ are two constants, $\gamma_c$ and $\gamma_m$ are
cooling Lorentz factor and minimum Lorentz factor of electrons
respectively (Sari et al. 1998). To compare with Fermi LAT
observations, we consider the electrons that produce $h\nu_*=100
{\rm MeV}$ synchrotron photons. We define $\gamma_*$ as the
Lorentz factor of those electrons whose synchrotron frequency is
$\nu_*$. The electrons producing such high-energy afterglow
emission typically have $\gamma_*\ga \max(\gamma_c,\gamma_m)$. For
the slow-cooling case, one can derive the number density of
electrons of $\gamma_*$ when  the KN effect is taken into account,
i.e.
\begin{equation}
N(\gamma_*)=\frac{1+Y(\gamma_c)}{1+Y(\gamma_*)}C_1\gamma_c
\gamma_*^{-p-1}=\frac{N_{syn}(\gamma_*)}{1+Y(\gamma_*)},
\end{equation}
where
$N_{syn}(\gamma_*)=C_1\gamma_{c,syn}\gamma_*^{-p-1}=C_1\gamma_{c}[1+Y(\gamma_c)]\gamma_*^{-p-1}$
represents the number density of electrons of $\gamma_*$ when only
the synchrotron cooling is considered ($\gamma_{c,syn}$ is the
cooling Lorentz factor of electrons when only the synchrotron
cooling is considered; see, e.g. Sari et al. 1998). Therefore the
number density of electrons of $\gamma_*$ is  a factor of
$1+Y(\gamma_*)$ lower than that in the case that only the
synchrotron cooling is considered. As a result, the synchrotron
luminosity produced by $\gamma_*$ electrons is reduced by the same
factor correspondingly. In the fast-cooling case, when the IC
scatterings of electrons of $\gamma_e\la\gamma_m$ with synchrotron
photons are in the Thomson scattering regime (the case of
$\S${3.2.2}), one can also obtain
$N(\gamma_*)={N_{syn}(\gamma_*)}/[{1+Y(\gamma_*)}]$, so the the
synchrotron luminosity is also reduced by a factor of
$1+Y(\gamma_*)$.

\section{KN effect on the Compton parameters}
Now we derive $Y(\gamma_*)$. As the electron distribution is
different in the fast and slow cooling cases, we divide the
following  analysis into these two different cases.

\subsection{The slow-cooling case}
Whether the afterglow emission belongs to the slow-cooling or
fast-cooling case depends on shock microphysics parameters (i.e.
the magnetic field equipartition factor $\epsilon_B$ and electron
energy equipartition factor $\epsilon_e$)  and other parameters
such as the burst energy $E$ and the circumburst density $n$.
Among these parameters, the magnetic field equipartition factor
$\epsilon_B$ is the mostly poorly known. The circumburst density
$n$ depends on the burst environment and may range from $10^{-3}
{\rm cm^{-3}}$ to $10 \,{\rm cm^{-3}}$ (e.g. Kumar \& Panaitescu
2001).

The condition for slow-cooling is
\begin{equation}
n_{-1}^{1/2}\epsilon_{B,-5}< 400 [1+Y(\gamma_c)]^{-1} f_p^{-1}
\epsilon_{e,-1}^{-1}E_{54}^{-1/2}t_0^{1/2}(1+z)^{-1/2},
\end{equation}
where $f_p\equiv 6(p-2)/(p-1)$, $p$ is the power-law index of the
electron energy distribution ($p=2.2$ has been used in the
following calculations), $t_0$ is the time in units of $10^0$ s
(hereafter we use the cgs units and denotation $Q_x=Q/10^x $) and
$z$ is the burst redshift. The cooling Lorentz factor and the
minimum Lorentz factor of electrons in forward shocks are given by
\begin{equation}
\gamma_c=10^7\left[{1+Y(\gamma_c)}\right]^{-1}
E_{54}^{-3/8}n_{-1}^{-5/8}\epsilon_{B,-5}^{-1}t_0^{1/8}(1+z)^{-1/8}
\end{equation}
and
\begin{equation}
\gamma_m=2.5\times10^4
f_p\epsilon_{e,-1}E_{54}^{1/8}n_{-1}^{-1/8}t_0^{-3/8}(1+z)^{3/8},
\end{equation}
respectively, where $Y(\gamma_c)$ is the Compton parameter of the
electrons of energy $\gamma_c$. The cooling frequency and minimum
frequency of electrons corresponding to $\gamma_c$ and $\gamma_m$
are, respectively,
\begin{equation}
\nu_c=8\times10^{22}\left[{1+Y(\gamma_c)}\right]^{-2}\epsilon_{B,-5}^{-3/2}E_{54}^{-1/2}n_{-1}^{-1}t_0^{-1/2}(1+z)^{-1/2}
{\rm Hz}
\end{equation}
and
\begin{equation}
\nu_m=5\times10^{17}f_p^2\epsilon_{e,-1}^2
\epsilon_{B,-5}^{1/2}E_{54}^{1/2}t_0^{-3/2}(1+z)^{1/2} {\rm Hz}.
\end{equation}

In the slow-cooling case, the synchrotron luminosity is dominated
by $\gamma_c$ electrons and the ratio of the SSC luminosity to
synchrotron luminosity is approximately given by $Y(\gamma_c)$.
Depending on the location of $\nu_{KN}(\gamma_c)$,
$U_{ph}[\nu<\nu_{KN}(\gamma_c)]$ is proportional to
${\nu_{KN}(\gamma_c)}^{(3-p)/2}$ or ${\nu_{KN}(\gamma_c)}^{4/3}$.
So the value of $Y(\gamma_c)$ can be obtained from
\begin{equation}
\begin{array}{lll}
Y(\gamma_c)[1+Y(\gamma_c)]=\frac{\epsilon_e}{\epsilon_B}(\frac{\gamma_c}{\gamma_m})^{2-p}
\\ \times \left \{
{\begin{array}{ll}
\left(\frac{\nu_m}{\nu_c}\right)^{(3-p)/2}\left(\frac{\nu_{KN}(\gamma_c)}{\nu_m}\right)^{4/3},
\,\,\,\,\, \nu_{KN}(\gamma_c)< \nu_m
\\\left(\frac{\nu_{KN}(\gamma_c)}{\nu_c}\right)^{(3-p)/2},
\,\,\,\,\,\,\,\,\,\,\,\,\,\,\,\,\,\,\,\,\,\,\,\,\,\,\,\,\,\,\,\,\,\,
\nu_m< \nu_{KN}(\gamma_c)<\nu_c
\\
1.\,\,\,\,\,\,\,\,\,\,\,\,\,\,
\,\,\,\,\,\,\,\,\,\,\,\,\,\,\,\,\,\,\,\,\,\,\,\,\,\,\,\,\,\,\,\,\,\,\,\,\,\,\,\,\,\,\,\,\,\,\,\,\,\,\,\,\,\,\,\,\,\,\,\,
\,\,\,\,\,\,\,\,\,\,\,\,\,\,\,\,\,\,\,\, \nu_c<\nu_{KN}(\gamma_c)
\end{array} }\right.
\end{array}
\end{equation}

To calculate $Y(\gamma_c)$, we need to know the ratios of
$\nu_{KN}(\gamma_c)$ to $\nu_m$ and $\nu_c$, which are
respectively given by
\begin{equation}
\frac{\nu_{KN}(\gamma_c)}{\nu_c}=7.5\times10^{-8}
\left[{1+Y(\gamma_c)}\right]^{3} E_{54}\epsilon_{B,-5}^{5/2}
n_{-1}^{3/2}(1+z)
\end{equation}
and
\begin{equation}
\frac{\nu_{KN}(\gamma_c)}{\nu_m}=1.4\times10^{-2}\left[{1+Y(\gamma_c)}\right]f_p^{-2}\epsilon_{e,-1}^{-2}\epsilon_{B,-5}^{1/2}n_{-1}^{1/2}
t_0 .
\end{equation}

Since we are interested in the high-energy afterglow emission, we
also need to know the ratio of $\nu_{KN}(\gamma_*)$ to $\nu_m$,
which is
\begin{equation}
\begin{array}{ll}
\frac{\nu_{KN}(\gamma_*)}{\nu_m}=(\frac{\nu_c}{\nu_*})^{1/2}\frac{\nu_{KN}(\gamma_c)}{\nu_m}
\\=2.8\times10^{-2}
f_p^{-2}\epsilon_{e,-1}^{-2}\epsilon_{B,-5}^{-1/4}E_{54}^{-1/4}
t_0^{3/4}(1+z)^{-1/4}.
\end{array}
\end{equation}

According to the relations among ${\nu_{KN}(\gamma_c)}$, $\nu_m$
and $\nu_c$,  we divide the discussion into three cases.
\subsubsection{Case I: $\nu_{KN}(\gamma_c)<{\nu_m}< \nu_c$}
This case typically happens at early times for reference parameter
values we used. Eq.(11) can be simplified as
\begin{equation}
\begin{array}{ll}
Y(\gamma_c)[1+Y(\gamma_c)]=0.09[1+Y(\gamma_c)]^{7/3} \\
\times
f_p^{-5/3}\epsilon_{e,-1}^{-2/3}\epsilon_{B,-5}^{2/3}E_{54}^{1/2}n_{-1}^{7/6}
t_0^{5/6}(1+z)^{1/2}.
\end{array}
\end{equation}
If $Y(\gamma_c)\ll 1$, one can obtain
\begin{equation}
Y(\gamma_c)=0.09
f_p^{-5/3}\epsilon_{e,-1}^{-2/3}\epsilon_{B,-5}^{2/3}E_{54}^{1/2}n_{-1}^{7/6}
t_0^{5/6}(1+z)^{1/2},
\end{equation}
while for $Y(\gamma_c)\ga 1$, the value of $Y(\gamma_c)$ can be
obtained only numerically.

One can also obtain the Compton parameter for those electrons that
produce high-energy synchrotron emission with frequency $\nu_*$.
Since usually $h\nu_*=100{\rm MeV}> h\nu_c$, we only discuss the
case of  $\nu_{KN}(\gamma_*)<\nu_{KN}(\gamma_c)$
below{\footnote{When the number density $n$ is very low (e.g.
$n\la 10^{-2}{\rm cm^{-2}}$), $h\nu_*$ can be lower than $h\nu_c$
and $\nu_{KN}(\gamma_*)>\nu_{KN}(\gamma_c)$.  $Y(\gamma_*)$ can be
similarly  obtained in this case. In figures 1 and 2 where
$Y(\gamma_*)$ is calculated for $n$ ranging from $10^{-3}$ to $10
{\rm cm^{-3}}$ and $\epsilon_B$ ranging from $10^{-1}$ to
$10^{-6}$, this situation has been included in the calculation.
}}. For $\nu_{KN}(\gamma_*)<\nu_{KN}(\gamma_c)<{\nu_m}$, we have
\begin{equation}
\begin{array}{ll}
Y(\gamma_*)=Y(\gamma_c)[\frac{\nu_{KN}(\gamma_*)}{\nu_{KN}(\gamma_c)}]^{4/3}=Y(\gamma_c)(\frac{\gamma_*}{\gamma_c})^{-4/3}=Y(\gamma_c) \left(\frac{\nu_*}{\nu_c}\right)^{-2/3} \\
=0.3f_p^{-5/3}\epsilon_{e,-1}^{-2/3}\epsilon_{B,-5}^{-1/3}E_{54}^{1/6}n_{-1}^{1/2}
t_0^{1/2}(1+z)^{1/6}.
\end{array}
\end{equation}

\subsubsection{  Case II: ${\nu_m}<{\nu_{KN}(\gamma_c)}<\nu_c$}
As ${\nu_{KN}(\gamma_c)}/{\nu_m}$ increases with time, it is
likely that ${\nu_{KN}(\gamma_c)}> {\nu_m}$ at later times. In
this case, we have
\begin{equation}
\begin{array}{ll}
Y(\gamma_c)[1+Y(\gamma_c)]=\frac{\epsilon_e}{\epsilon_B}(\frac{\gamma_c}{\gamma_m})^{2-p}(\frac{\nu_{KN}(\gamma_c)}{\nu_c})^{(3-p)/2}
\\
=1.2[1+Y(\gamma_c)]^{(5-p)/2}f_p^{p-2}\epsilon_{e,-1}^{p-1}\epsilon_{B,-5}^{(3-p)/4}E_{54}^{1/2}n_{-1}^{(5-p)/4}
t_1^{(2-p)/2}(1+z)^{1/2}.
\end{array}
\end{equation}
Depending on whether $\nu_{KN}(\gamma_*)$ is larger or smaller
than $\nu_m$, there are two sub-cases:   \\
1) Case IIa: $\nu_{KN}(\gamma_c)> {\nu_m}>{\nu_{KN}(\gamma_*)}$.
In this case,
\begin{equation}
\begin{array}{ll}
Y(\gamma_*)=Y(\gamma_c)[\frac{\nu_m}{\nu_{KN}(\gamma_c)}]^{(3-p)/2}[\frac{\nu_{KN}(\gamma_*)}{\nu_m}]^{4/3}\\
=0.9
f_p^{-5/3}\epsilon_{e,-1}^{-2/3}\epsilon_{B,-5}^{-1/3}E_{54}^{1/6}n_{-1}^{1/2}
t_1^{1/2}(1+z)^{1/6}.
\end{array}
\end{equation}
\\2) Case IIb: $\nu_{KN}(\gamma_c)>{\nu_{KN}(\gamma_*)}>
{\nu_m}$. In this case,
\begin{equation}
\begin{array}{ll}
Y(\gamma_*)=Y(\gamma_c)[\frac{\nu_{KN}(\gamma_*)}{\nu_{KN}(\gamma_c)}]^{(3-p)/2}
=Y(\gamma_c) (\frac{\nu_*}{\nu_{c}})^{(p-3)/4}
\\=2f_p^{p-2}\epsilon_{e,-1}^{p-1}\epsilon_{B,-5}^{(p-3)/8}E_{54}^{(p+1)/8}n_{-1}^{1/2}
t_2^{(5-3p)/8}(1+z)^{(p-2)/2}.
\end{array}
\end{equation}

\subsubsection{Case III: $\nu_{KN}(\gamma_c)>\nu_c>\nu_m$}
In this case, the KN effect on $\gamma_c$ electrons is not
important and the Compton parameter is given by
\begin{equation}
\begin{array}{ll}
Y(\gamma_c)[1+Y(\gamma_c)]=\frac{\epsilon_e}{\epsilon_B}(\frac{\gamma_c}{\gamma_m})^{2-p}=2.4\times10^3[1+Y(\gamma_c)]^{p-2}f_p^{p-2}\\
\times\epsilon_{e,-1}^{p-1}\epsilon_{B,-5}^{p-3}E_{54}^{(p-2)/2}n_{-1}^{(p-2)/2}
t_1^{-(p-2)/2}(1+z)^{(p-2)/2}.
\end{array}
\end{equation}
Depending on the relations among $\nu_{KN}(\gamma_*)$, $\nu_m$ and
$\nu_c$, there are three sub-cases:
\\ 1) Case III a: $\nu_{KN}(\gamma_*)<\nu_m<\nu_c<\nu_{KN}(\gamma_c)$. Define $\hat\gamma_c=\Gamma m_e c^2/\nu_c$ as the critic Lorentz
factor of those electrons that their interaction with synchrotron
peak photons (i.e. photons at $\nu_c$) is just in the KN regime.
In this case, we have
\begin{equation}
\begin{array}{lll}
Y(\gamma_*)=Y(\hat\gamma_c)(\frac{\nu_m}{\nu_c})^{(3-p)/2}[\frac{\nu_{KN}(\gamma_*)}{\nu_m}]^{4/3}\\
=0.9f_p^{-5/3}\epsilon_{e,-1}^{-2/3}\epsilon_{B,-5}^{-1/3}E_{54}^{1/6}n_{-1}^{1/2}
t_1^{1/2}(1+z)^{1/6},
\end{array}
\end{equation}
where $Y(\hat\gamma_c)=Y(\gamma_c)$ has been used in the last
step.
\\ 2) Case IIIb:
$\nu_m<\nu_{KN}(\gamma_*)<\nu_c<\nu_{KN}(\gamma_c)$. In this case,
\begin{equation}
\begin{array}{ll}
Y(\gamma_*)=Y(\hat\gamma_c)[\frac{\nu_{KN}(\gamma_*)}{\nu_c}]^{(3-p)/2}=Y(\gamma_c)[\frac{\nu_*}{\nu_c}]^{(p-3)/4}
(\frac{\nu_{KN}(\gamma_c)}{\nu_c})^{(3-p)/2}\\
=2f_p^{p-2}\epsilon_{e,-1}^{p-1}\epsilon_{B,-5}^{(p-3)/8}E_{54}^{(p+1)/8}n_{-1}^{1/2}
t_2^{(5-3p)/8}(1+z)^{(p-2)/2}.
\end{array}
\end{equation}
\\ 3) Case IIIc:
$\nu_m<\nu_c<\nu_{KN}(\gamma_*)<\nu_{KN}(\gamma_c)$. In this case,
\begin{equation}
Y(\gamma_*)=Y(\gamma_c) .
\end{equation}

\subsection{The fast-cooling case} The condition for the
fast-cooling case is
\begin{equation}
n_{-1}^{1/2}\epsilon_{B,-2}\ga 0.4 [1+Y(\gamma_c)]^{-1} f_p^{-1}
\epsilon_{e,-1}^{-1}E_{54}^{-1/2}t_0^{1/2}(1+z)^{-1/2}.
\end{equation}
Below we use a larger $\epsilon_B$  as the reference value for the
fast-cooling case. In this case, the synchrotron radiation is
dominated by $\gamma_m$ electrons and the critic frequency of
interest is
\begin{equation}
\nu_{KN}(\gamma_m)=3.7\times10^{18}f_p^{-1}\epsilon_{e,-1}^{-1}
 {\rm Hz}.
\end{equation}
Similarly, one can find the ratios of $\nu_{KN}(\gamma_m)$ to two
characteristic frequencies,
\begin{equation}
\frac{\nu_{KN}(\gamma_m)}{\nu_m}=0.24f_p^{-3}\epsilon_{e,-1}^{-3}
E_{54}^{-1/2}\epsilon_{B,-2}^{-1/2}t_0^{3/2}(1+z)^{-1/2},
\end{equation}
\begin{equation}
\frac{\nu_{KN}(\gamma_m)}{\nu_c}=1.5\left[{1+Y(\gamma_c)}\right]^{2}f_p^{-1}\epsilon_{e,-1}^{-1}
E_{54}^{1/2}\epsilon_{B,-2}^{3/2}n_{-1}t_0^{1/2}(1+z)^{1/2}.
\end{equation}

Below we divide the discussion into two cases according to whether
the KN effect of $\gamma_m$ electrons is important or not, i.e.
the cases of $\nu_{KN}(\gamma_m)<\nu_m$ and
$\nu_{KN}(\gamma_m)>\nu_m$.
\subsubsection {Case I: $\nu_{KN}(\gamma_m)<\nu_m$}
Different from the slow-cooling case, the electron distribution at
low-energies is affected by the KN effect in this case and
therefore the corresponding synchrotron spectrum may be changed.
Following Nakar et al. (2009), we define $\nu_0$ as the
synchrotron frequency of electrons of $\gamma_0$ (i.e.
$\nu_0=\nu_{\rm syn} {(\gamma_0)}$), where $Y(\gamma_0)=1$.
According to whether $\gamma_m$ is greater or smaller than
$\gamma_0$, there are two subcases, i.e. 1) $\gamma_0< \gamma_m$
and 2)$\gamma_0> \gamma_m$.

i)Case Ia: $\gamma_0<\gamma_m$. This case applies when
$\epsilon_B$ is large. Define $\hat\gamma_0=\Gamma m_e c^2/h\nu_0$
and $\hat\gamma_m=\Gamma m_e c^2/h\nu_m$. In the energy range
$\hat\gamma_0<\gamma_e<\gamma_0$, the electron distribution is
$N(\gamma_e)\propto \gamma_e^{-1}$ and the synchrotron spectrum is
$\nu F_\nu\propto \nu$ (Wang et al. 2009; Nakar et al. 2009).
Hence we have
$Y(\gamma_0)=Y(\hat\gamma_0)(\gamma_0/\hat\gamma_0)^{-1}$. Since
$\nu F_{\nu}(\nu_0)=\nu F_{\nu}(\nu_m)(\nu_0/\nu_m)^{1/2}$ and
$Y(\hat\gamma_m)=\epsilon_e/\epsilon_B$ in this case,
$Y(\hat\gamma_0)=\frac{\epsilon_e}{\epsilon_B}(\frac{\nu_0}{\nu_m})^{1/2}$.
Then we obtain
$\gamma_0=\hat\gamma_0(\frac{\epsilon_e}{\epsilon_B})(\frac{\nu_0}{\nu_m})^{1/2}$.
From
\begin{equation}
\frac{\nu_0}{\nu_m}=(\frac{\gamma_0}{\gamma_m})^2=0.06f_p^{-3}\epsilon_{e,-1}^{-2}
E_{54}^{-1/2}\epsilon_{B,-1}^{-3/2}t_0^{3/2}(1+z)^{-1/2},
\end{equation}
one can further obtain the corresponding synchrotron frequency of
$\gamma_0$ electrons,
\begin{equation}
\nu_0=3.6\times10^{18}f_p^{-1}\epsilon_{B,-1}^{-1} {\rm Hz} .
\end{equation}
Similarly, one can obtain the synchrotron frequency of
$\hat\gamma_0$ electrons,
\begin{equation}
\begin{array}{ll}
{\hat\nu_0}=(\frac{\hat\gamma_0}{\gamma_m})^2{\nu_m}=[\frac{\nu_{KN}(\gamma_m)}{\nu_0}]^2{\nu_m}\\=5\times10^{19}f_p^2
\epsilon_{B,-1}^{5/2}E_{54}^{1/2}t_0^{-3/2}(1+z)^{1/2} {\rm Hz}.
\end{array}
\end{equation}
If $\max(\nu_c,\hat\nu_0)\la\nu_{KN}(\gamma_m)\la \nu_0\la \nu_m$,
we have
\begin{equation}
Y(\gamma_m)[1+Y(\gamma_m)]=\frac{\epsilon_e}{\epsilon_B}(\frac{\nu_{KN}(\gamma_m)}{\nu_0})(\frac{\nu_0}{\nu_m})^{1/2}.
\end{equation}
As $Y(\gamma_m)<1$ when $\gamma_0<\gamma_m$, we obtain
\begin{equation}
\begin{array}{ll}
Y(\gamma_m)\simeq\frac{\epsilon_e}{\epsilon_B}(\frac{\nu_{KN}(\gamma_m)}{\nu_0})(\frac{\nu_0}{\nu_m})^{1/2}
=0.2f_p^{-3/2}\epsilon_{e,-1}^{-1}
E_{54}^{-1/4}\epsilon_{B,-1}^{-3/4}t_0^{3/4}(1+z)^{-1/4}.
\end{array}
\end{equation}
If
$\max(\nu_c,\hat\nu_0)\la\nu_{KN}(\gamma_*)\la\nu_{KN}(\gamma_m)\la\nu_0$,
it is easy to obtain
\begin{equation}
Y(\gamma_*)=Y(\gamma_m)\frac{\nu_{KN}(\gamma_*)}{\nu_{KN}(\gamma_m)}=Y(\gamma_m)(\frac{\nu_*}{\nu_m})^{-1/2}=
0.01f_p^{-1/2}\epsilon_{B,-1}^{-1/2}.
\end{equation}
In other cases, the derivation of $Y(\gamma_*)$ is complicated.
However, we note that as long as
$\nu_*>\nu_0=3.6\times10^{18}f_p^{-1}\epsilon_{B,-1}^{-1} {\rm Hz}
$,
\begin{equation}
Y(\gamma_*)<1.
\end{equation}
For $h\nu_*=100 {\rm MeV}$, $\nu_*>\nu_0$ is satisfied given that
the fast-cooling condition is satisfied. Therefore we conclude
that $Y(\gamma_*)<1$ in case Ia.

ii) Case Ib: $\gamma_0>\gamma_m$. This case applies when
$\epsilon_B$ is smaller. In this case, $\nu_{KN}(\gamma_m)\la
\nu_m\la \nu_0$, so
\begin{equation}
Y(\gamma_m)[1+Y(\gamma_m)]=\frac{\epsilon_e}{\epsilon_B}\frac{\nu_{KN}(\gamma_m)}{\nu_m}.
\end{equation}
From $Y(\gamma_0)=Y(\gamma_m)(\frac{\gamma_0}{\gamma_m})^{-1}=1$,
we obtain
$\gamma_0=\frac{\epsilon_e}{\epsilon_B}\frac{\nu_{KN}(\gamma_m)}{\nu_m}\gamma_m$
and
\begin{equation}
\nu_0=3.6\times10^{19}f_p^{-1}\epsilon_{B,-2}^{-1} {\rm Hz}.
\end{equation}
As, $Y(\gamma_m)>1$ in this case, we get
\begin{equation}
\begin{array}{ll}
Y(\gamma_m)=\left[\frac{\epsilon_e}{\epsilon_B}\frac{\nu_{KN}(\gamma_m)}{\nu_m}\right]^{1/2}
=1.2f_p^{-3/2}\epsilon_{e,-1}^{-1}
E_{54}^{-1/4}\epsilon_{B,-2}^{-3/4}t_0^{3/4}(1+z)^{-1/4}.
\end{array}
\end{equation}
Similarly, if
$\max(\nu_c,\hat\nu_0)\la\nu_{KN}(\gamma_*)\la\nu_{KN}(\gamma_m)\la\nu_0$,
\begin{equation}
Y(\gamma_*)=Y(\gamma_m)\frac{\nu_{KN}(\gamma_*)}{\nu_{KN}(\gamma_m)}=
0.03f_p^{-1/2}\epsilon_{B,-2}^{-1/2}
\end{equation}
In other case, we also have
\begin{equation}
Y(\gamma_*)<1
\end{equation}
as long as $\nu_*>\nu_0$ is satisfied.

\subsubsection{Case II: $\nu_{KN}(\gamma_m)>\nu_m$}
At later times, when $t> 3 f_p^2\epsilon_{e,-1}^2
E_{54}^{1/3}\epsilon_{B,-2}^{1/3}(1+z)^{1/3} {\rm s}$,
$\nu_{KN}(\gamma_m)> \nu_m$. In this case, the KN effect of
$\gamma_m$ electrons is unimportant and
\begin{equation}
Y(\gamma_m)=(\frac{\epsilon_e}{\epsilon_B})^{1/2}=3
\epsilon_{e,-1}^{1/2}\epsilon_{B,-2}^{-1/2},
\end{equation}
as the synchrotron emission typically peaks at $\nu_m$ in this
case (Nakar et al. 2009). In order to calculate $Y(\gamma_*)$,
let's first derive the ratios of $\nu_{KN}(\gamma_*)$ to two
critic frequencies, which are respectively
\begin{equation}
\begin{array}{ll}
\frac{\nu_{KN}(\gamma_*)}{\nu_m}=(\frac{\nu_m}{\nu_*})^{1/2}\frac{\nu_{KN}(\gamma_m)}{\nu_m}\\=0.04
f_p^{-2}\epsilon_{e,-1}^{-2}
\epsilon_{B,-2}^{-1/4}E_{54}^{-1/4}t_1^{3/4}(1+z)^{-1/4}
\end{array}
\end{equation}
and
\begin{equation}
\begin{array}{ll}
\frac{\nu_{KN}(\gamma_*)}{\nu_c}=(\frac{\nu_m}{\nu_*})^{1/2}\frac{\nu_{KN}(\gamma_m)}{\nu_c}\\=0.25\epsilon_{e,-1}\epsilon_{B,-2}^{3/4}
E_{54}^{3/4}n_{-1}t_1^{-1/4}(1+z)^{3/4},
\end{array}
\end{equation}
where $Y(\gamma_c)=Y(\gamma_m)=\sqrt{\epsilon_e/\epsilon_B}$ has
been used. According to the relations among $\nu_{KN}(\gamma_*)$,
$\nu_m$ and $\nu_c$, there are three subcases:
\\1)Case IIa:  $\nu_c<\nu_{KN}(\gamma_*)<\nu_m$. In this
case,
\begin{equation}
Y(\gamma_*)=Y(\hat\gamma_m)(\frac{\nu_{KN}(\gamma_*)}{\nu_m})^{1/2}
=0.7f_p^{-1}\epsilon_{e,-1}^{-1/2}
\epsilon_{B,-2}^{-5/8}E_{54}^{-1/8}t_1^{3/8}(1+z)^{-1/8},
\end{equation}
where $\hat\gamma_m=\Gamma m_e c^2/h\nu_m$ is the critic Lorentz
factor of those electrons that their interaction with photons at
$\nu_m$ is just in the KN regime and
$Y(\hat\gamma_m)=Y(\gamma_m)=\sqrt{\epsilon_e/\epsilon_B}$ has
been used in the last step.
\\ 2)Case IIb: $\nu_{KN}(\gamma_*)<\nu_c$. In this case,
\begin{equation}
\begin{array}{ll}
Y(\gamma_*)=Y(\hat\gamma_m)(\frac{\nu_c}{\nu_m})^{1/2}(\frac{\nu_{KN}(\gamma_*)}{\nu_c})^{4/3}
=2.2 [1+Y(\gamma_c)]^{-1} f_p^{-1}\epsilon_{e,-1}^{5/6}\\ \times
\epsilon_{B,-2}^{-1/2}E_{54}^{1/2}n_{-1}^{5/6}t_1^{1/6}(1+z)^{1/2}
=0.18f_p^{-1}\epsilon_{e,-1}^{1/3}E_{54}^{1/2}n_{-1}^{5/6}t_1^{1/6}(1+z)^{1/2},
\end{array}
\end{equation}
\\ 3)Case IIc: $\nu_{KN}(\gamma_*)>\nu_m$. In this case,
\begin{equation}
Y(\gamma_*)=Y(\gamma_m)=3
\epsilon_{e,-1}^{1/2}\epsilon_{B,-2}^{-1/2}.
\end{equation}

\section{KN effect on the high-energy synchrotron afterglow luminosity }
\begin{figure}
\epsscale{2.5} \plottwo{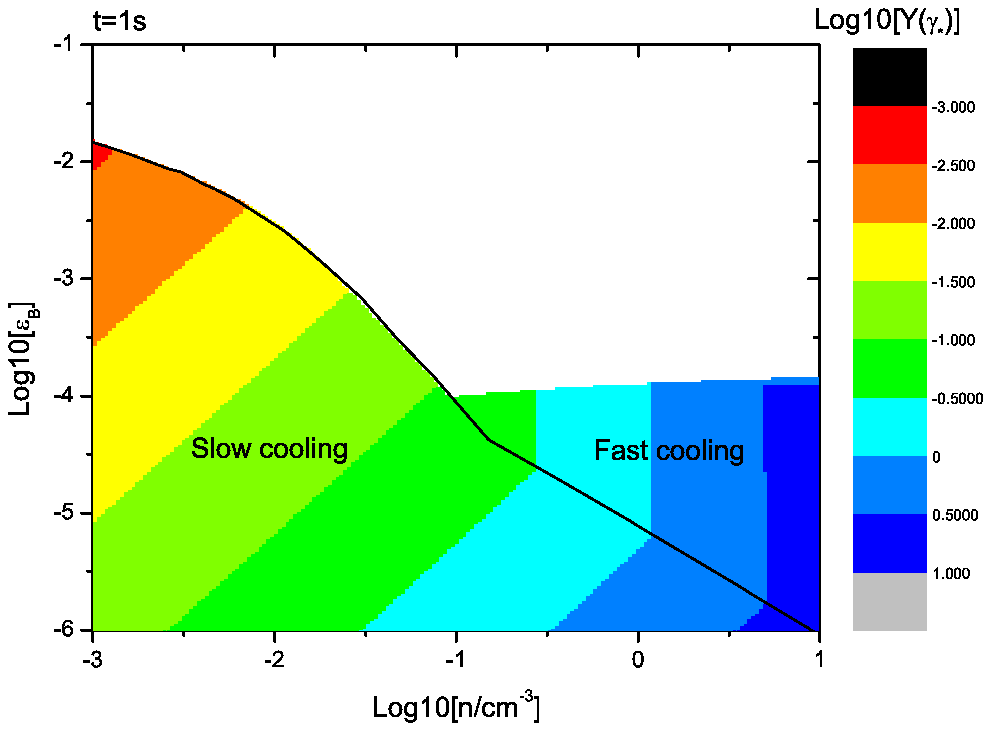}{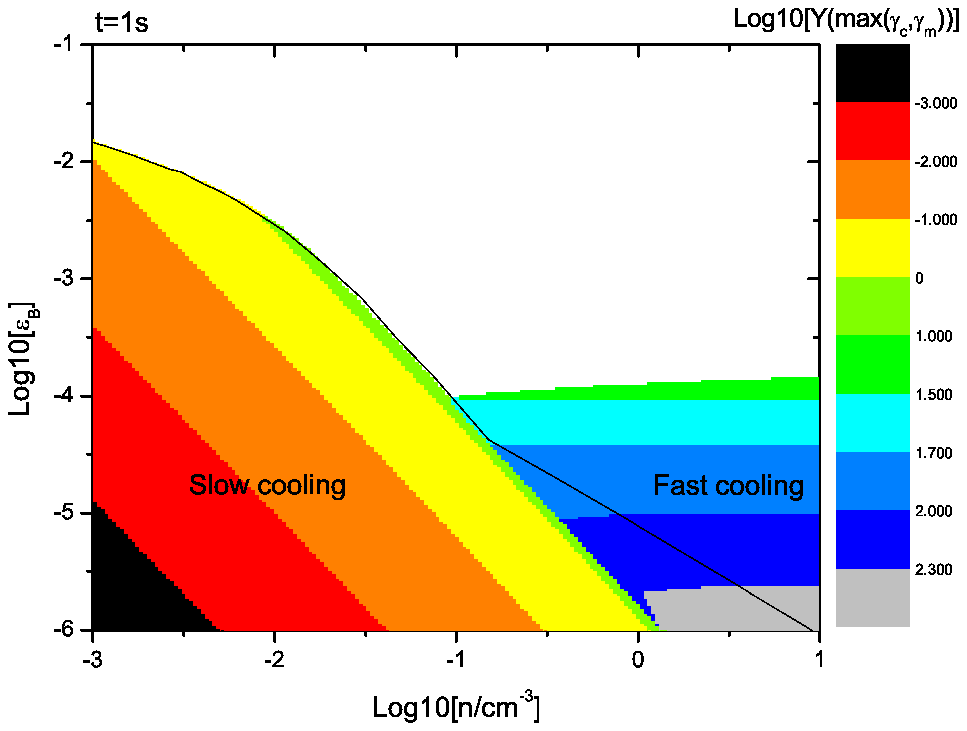}\caption{Values
of the Compton parameters $Y(\gamma_*)$ (top panel) and
$Y(\gamma_c)$ (or $Y(\gamma_m)$, bottom panel)  of the afterglow
emission as a function of $\epsilon_B$ and $n$ at time $t=1$s
after the burst. The black solid line separates the slow-cooling
case and fast-cooling case. The blank space in the plot
corresponds to the region where $\nu_{KN}(\gamma_m)<\nu_m$, for
which $Y(\gamma_*)<1$ but the exact value is not calculated. Other
parameters used in the plots are $\epsilon_e=0.1$ and
$E=10^{54}{\rm erg}$.}
\end{figure}

\begin{figure}
\epsscale{2.5} \plottwo{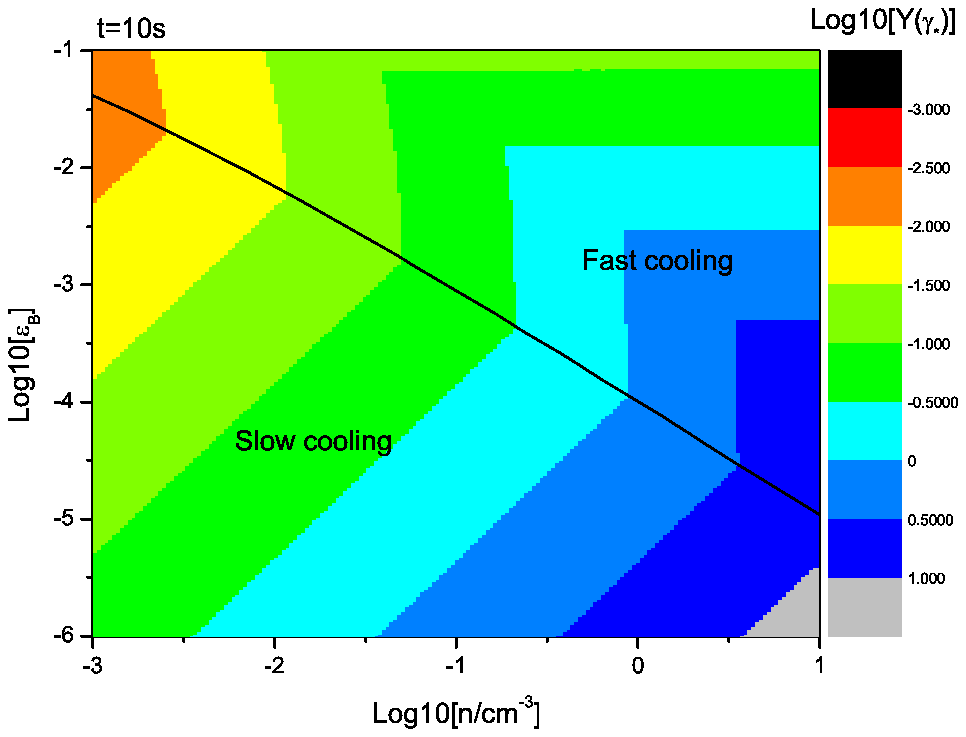}{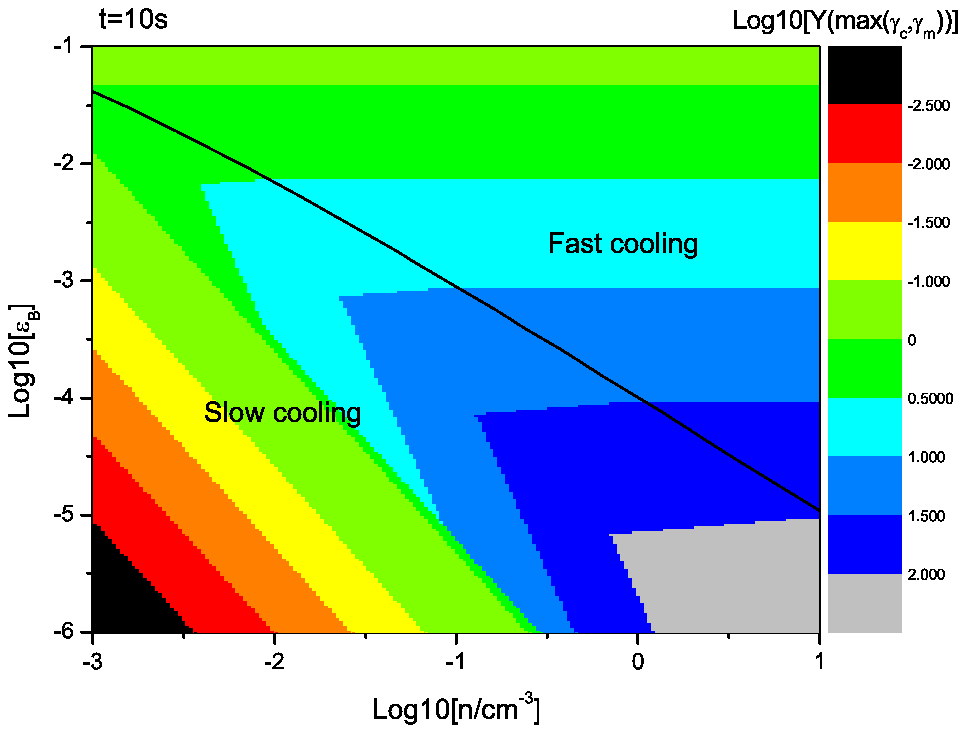}\caption{The
same as figure 1, but at time $t=10$ s after the burst.}
\end{figure}
The above analyses give the dependence of the Compton parameters
$Y(\gamma_*)$, $Y(\gamma_c)$ (in the slow-cooling case) and
$Y(\gamma_m)$ (in the fast-cooling case) on the parameters such as
$\epsilon_e$, $\epsilon_B$, $E$ and $n$. Since $\epsilon_B$ and
$n$ are the least known among these parameters for GRB afterglows,
we explore the value of $Y(\gamma_*)$ and $Y(\gamma_c)$ (or
$Y(\gamma_m)$) as a function of these two parameters. In Figures 1
and 2, we show the result for two different times, i.e. at $t=1
{\rm s}$ and $t=10 {\rm s}$ respectively. We find  $Y(\gamma_*)$
is smaller than a few at $t=1$ s for the parameters $\epsilon_B$
in the range from $10^{-6}$ to $10^{-1}$ and $n$ in the range from
$10^{-3} {\rm cm^{-3}}$ to $10 {\rm cm^{-3}}$. At $t=10$ s,
$Y(\gamma_*)$ is also smaller than a few in a wide range of
parameter space (it is large than a few only when $n$ is as high
as $10 {\rm cm^{-3}}$ and $\epsilon_B$ is close to $10^{-6}$). On
the other hand, $Y(\gamma_c)$ or $Y(\gamma_m)$ can be more than
one order of magnitude higher in the same parameter space. This
implies that SSC loss of high-energy electrons that produce
high-energy ($\ga 100$ MeV) afterglow photons is typically small.
As a result, the synchrotron luminosity at high-energies is
correspondingly high, which enables the detection of early
high-energy afterglow emission by Fermi LAT.

\begin{figure}
\plotone{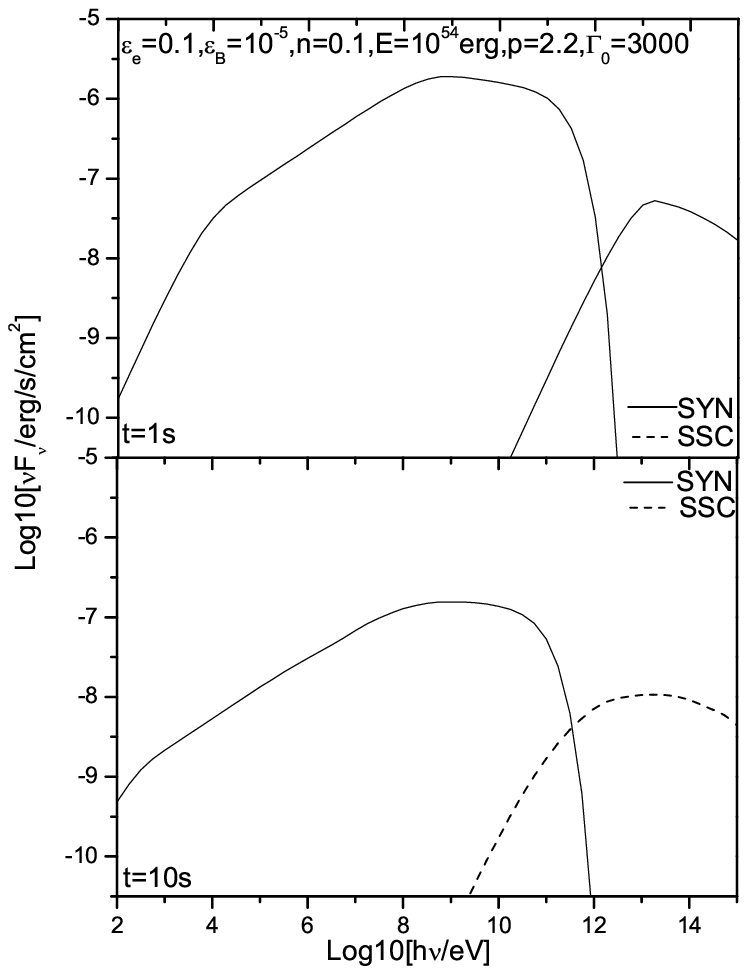} \caption{The calculated spectral
energy distribution of the afterglow emission in the slow-cooling
case at times $t=1$ s and $t=10$ s after the burst with the KN
effect  taken into account.  The solid and dashed lines represent
the synchrotron component and SSC component, respectively. }
\end{figure}
\begin{figure}
\plotone{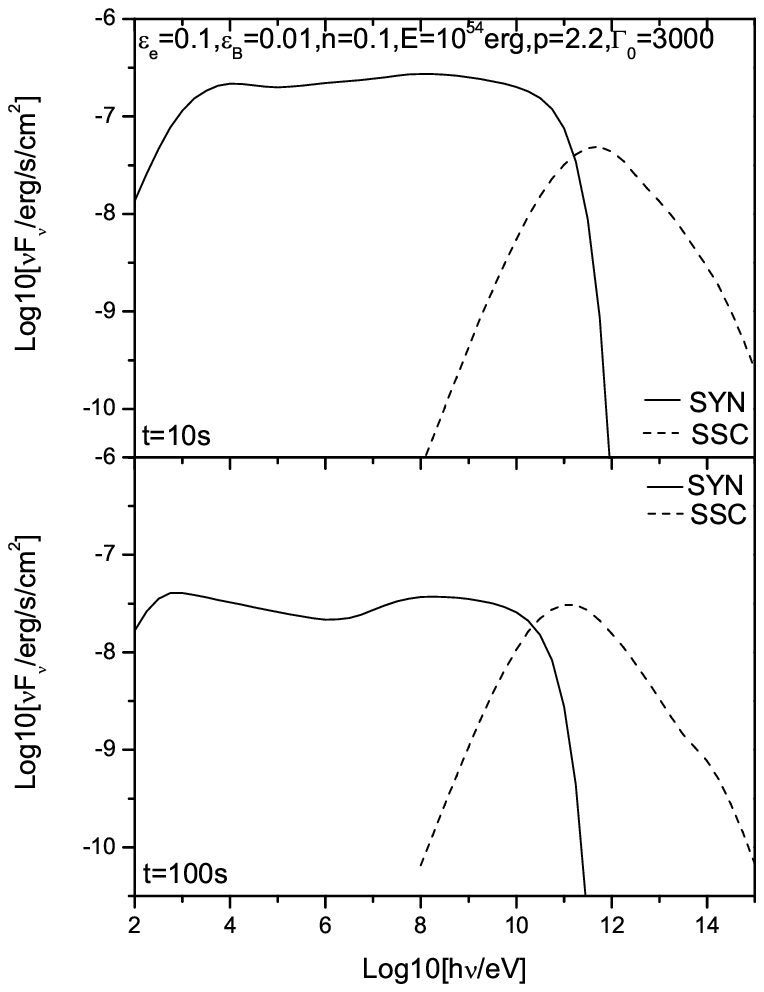} \caption{The calculated spectral
energy distribution of the afterglow emission in the fast-cooling
case at times $t=10$ s and $t=100$ s after the burst with the KN
effect taken into account.  The solid and dashed lines represent
the synchrotron component and SSC component, respectively. }
\end{figure}

\begin{figure}
\plotone{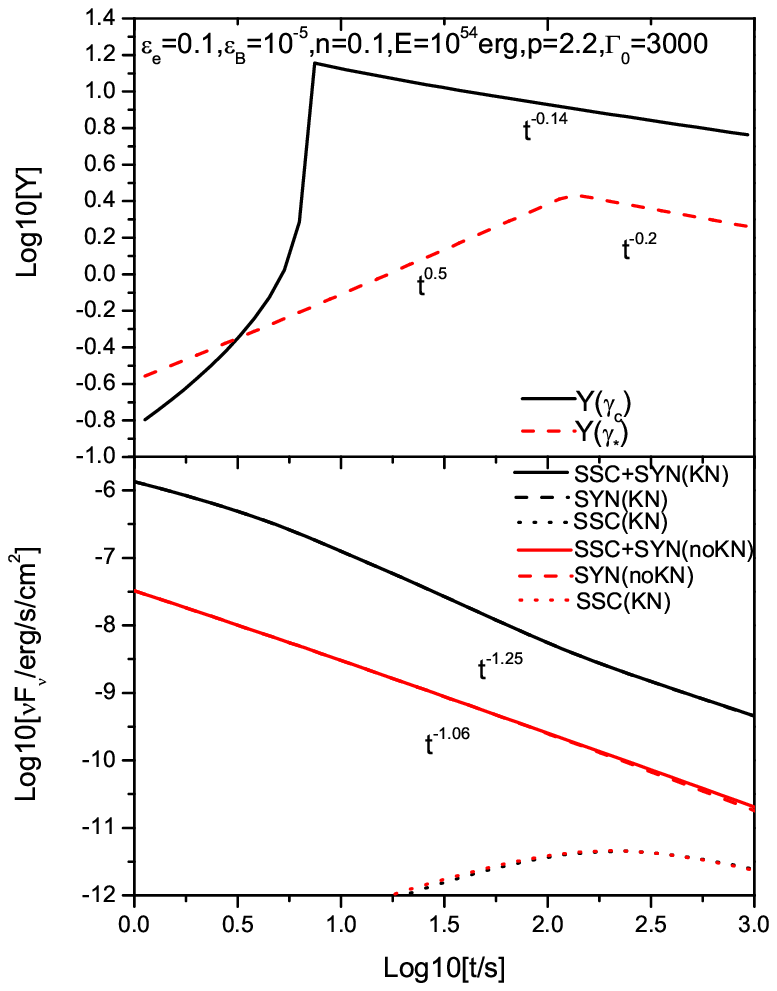} \caption{The top panel shows the
evolution of Compton parameters $Y(\gamma_*)$ and $Y(\gamma_c)$
with time in the slow-cooling case for the parameters
$\epsilon_e=0.1$, $\epsilon_B=10^{-5}$, $n=0.1{\rm cm^{-3}}$,
$E=10^{54}{\rm ergs}$, $p=2.2$ and $\Gamma_0=3000$. The bottom
panel shows the light curves of the synchrotron emission, the SSC
emission and the sum of them at $h\nu_*=100$ MeV. Black lines and
red lines denote, respectively, the light curves with and without
KN effect taken into account.}
\end{figure}

\begin{figure}
\plotone{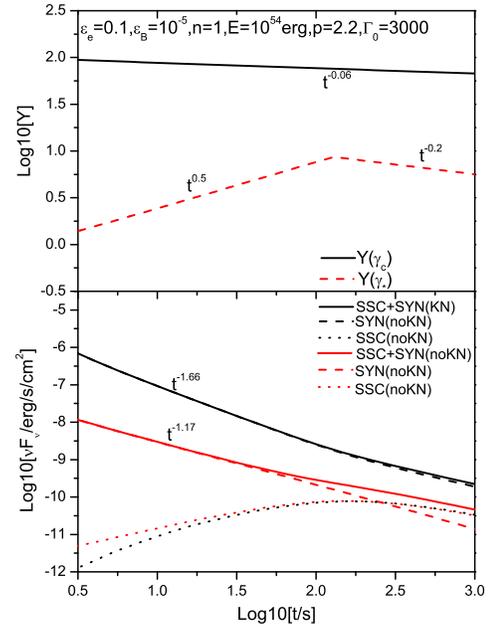} \caption{The same as figure 5,
but a larger circumburst density of $n=1 {\rm cm^{-3}}$.}
\end{figure}
In order to see whether the SSC emission contributes to the
high-energy afterglow emission at Fermi LAT energy band, we
calculate the spectral energy distribution of the afterglow
emission at early times numerically. Assuming an adiabatic
evolution of the blast wave and using the electron distribution
given in $\S$2, we calculate the synchrotron radiation spectrum as
well as the SSC spectrum with a full KN cross section taken into
account (see equations 2 and 11 of He et al. (2009) for the
description of the dynamic and the full KN cross section). Figure
3 shows the $\nu F_\nu$ spectra of the afterglow synchrotron
emission and the SSC emission for the slow-cooling case at times
$t=1$ s and $t=10$ s when the KN effect is taken into account. In
the Fermi LAT energy band, the synchrotron component is dominated
at both times. The SSC component becomes dominated only at
energies above the maximum synchrotron photon energy of
shock-accelerated electrons, at which the flux usually becomes,
however, too low to be detectable by Fermi LAT. The spectral
energy distribution of the afterglow emission for the fast-cooling
case at times $t=10$ s and $t=100$ s is shown in Fig.4. Similarly,
SSC contribution to the high-energy emission at energies below 100
GeV is negligible at these times. Fig.4 (see the bottom panel)
also shows that  the spectrum becomes harder at energies above
$10^7$ eV. This is caused by the decreased IC loss suppression on
the synchrotron flux at high energies due to the KN effect.

The Compton parameters also vary with time. In the above analytic
calculation in $\S$3,  we have shown that, $Y(\gamma_*)$ increases
with time as $t^{1/2}$  in the slow-cooling case as long as
$\nu_{KN}(\gamma_*)<\nu_m$. When $\nu_{KN}(\gamma_*)>\nu_m$,
$Y(\gamma_*)$ starts to decrease with time. We calculate
$Y(\gamma_c)$  numerically using Eq.11 and show the evolution of
$Y(\gamma_c)$ and $Y(\gamma_*)$ with time in the top panel of
Fig.5.  If $Y(\gamma_*)\ga1$ as well, as in the case of some
parameter space shown in figures 1 and 2, the decay of the
synchrotron afterglow emission will be faster than what is
predicted by the standard synchrotron afterglow theory (i.e.
steeper than $t^{(2-3p)/4}$ for $\nu_*>\nu_c$), since the
synchrotron luminosity at frequency $\nu_*$ scales as
$1/[1+Y(\gamma_*)]$. The decay could be steeper by a factor
$\Delta \alpha=1/2$ at most. We numerically calculate the light
curves of the afterglow emission assuming an adiabatic evolution
of the blast wave   and using the electron distribution given in
$\S$ 2. The light curves of the synchrotron emission, the SSC
emission and the sum of them at frequency $\nu_*=100 {\rm MeV}$
are shown in the bottom panel of figure 5 (the black lines). As a
comparison, we also show the light curves (the red lines) in the
case that the KN effect is not taken into account (i.e. assuming
that the SSC cooling is in the Thomson regime). One can see that
the light curve decay becomes steeper and the flux at 100 MeV is
significantly higher when the KN effect is taken into account. In
Fig.6, we also show the light curves for another set of parameters
in the slow-cooling case. It also shows that the temporal decay of
high-energy afterglow emission can be significantly steeper than
what is predicted by the standard synchrotron theory when the KN
effect is taken into account.

\begin{figure}
\plotone{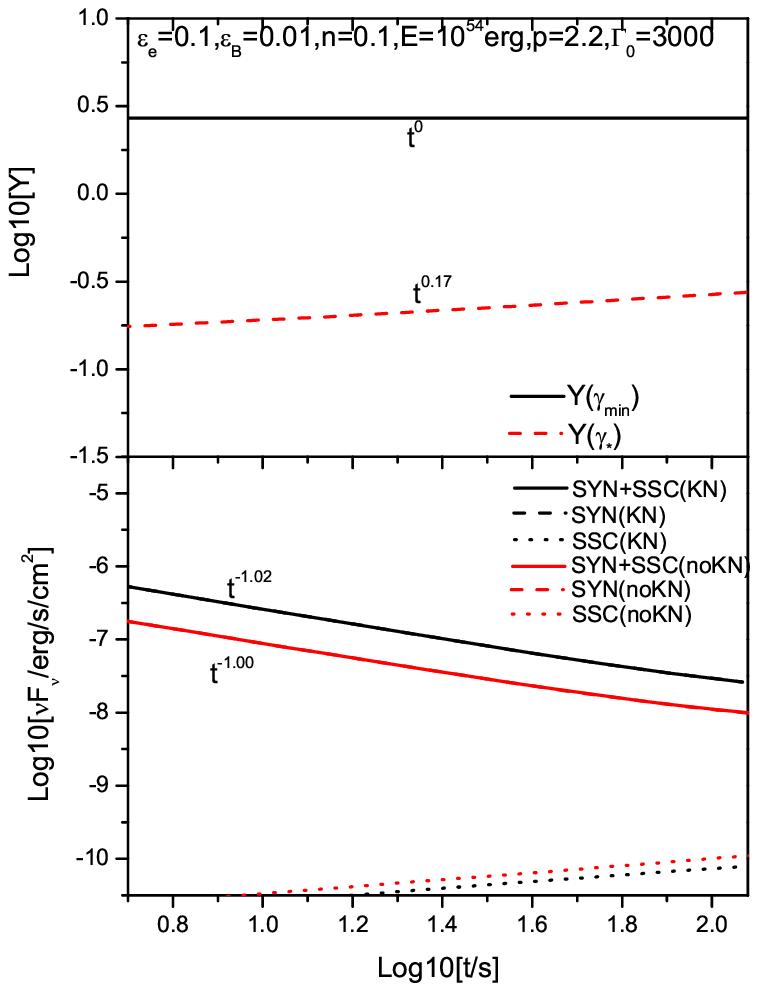}\caption{The top panel shows the
evolution of Compton parameters $Y(\gamma_*)$ and $Y(\gamma_m)$
with time in the fast-cooling case for the parameters
$\epsilon_e=0.1$, $\epsilon_B=0.01$, $n=0.1{\rm cm^{-3}}$,
$E=10^{54}{\rm ergs}$, $p=2.2$ and $\Gamma_0=3000$. The bottom
panel shows the light curves of the synchrotron emission, the SSC
emission and the sum of them at $h\nu_*=100$ MeV. Black lines and
red lines denote, respectively, the light curves with and without
KN effect taken into account.}
\end{figure}
\begin{figure}
\plotone{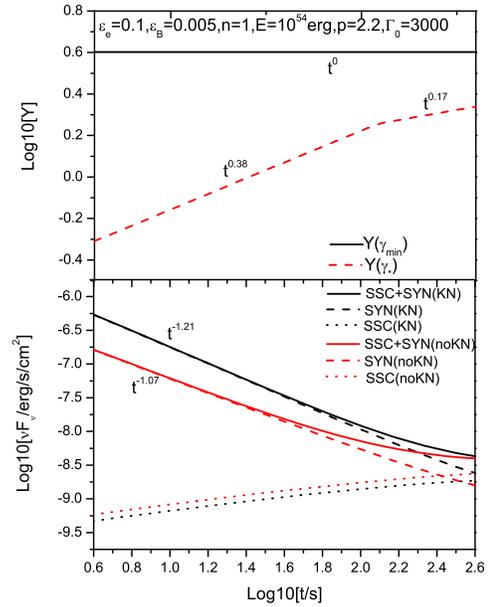} \caption{The same as figure 7,
but for $\epsilon_B=0.005$ and $n=1 {\rm cm^{-3}}$.}
\end{figure}

For the fast-cooling case, in the above analytic calculations we
have found that when $\nu_{KN}(\gamma_m)<\nu_m$, $\nu_*$ is always
larger than $\nu_0$ (except when $\epsilon_B<10^{-5}$, but note
that for $\epsilon_B<10^{-5}$, the afterglow emission is no longer
in the fast-cooling case and the discussion in $\S$3.1 applies).
For $\nu_*>\nu_0$, $Y(\gamma_*)<1$, so almost all of the energy of
high-energy electrons goes into the synchrotron emission and the
synchrotron luminosity is high.  As $Y(\gamma_*)<1$, the decay
slope of the high-energy emission will not be affected. However,
in the case of $\nu_{KN}(\gamma_m)>\nu_m$, $Y(\gamma_*)$ could be
larger than 1 and it increases with time as $t^{3/8}$ or $t^{1/6}$
in some certain parameter space. This will lead to a steeper decay
of the synchrotron afterglow emission at frequency $\nu_*$.
Figures 7 and 8 show the time  evolution of the Compton parameters
$Y(\gamma_*)$ and $Y(\gamma_m)$, and the light curves of the
high-energy afterglow emission at  $\nu_*=100 {\rm MeV}$ for two
set of parameters. They clearly indicate that the temporal decay
of high-energy afterglow emission becomes  steeper than
$t^{(2-3p)/4}$ when $Y(\gamma_*)\ga 1$. So we conclude that the
light curve of the high-energy afterglow emission could be also
steeper in the fast-cooling case when the KN effect is taken into
account.

\section{Implication for Fermi LAT observation of afterglow emission}
As has been shown in figures 1 and 2, the Compton parameters for
the electrons that produce high-energy gamma-ray afterglow
emission are typically small at early time $t\la 10 {\rm s}$, i.e.
$Y(\gamma_*)\la {\rm a ~few}$ for a wide range of parameter space.
This has important implication for the detectability of
high-energy afterglow emission by Fermi LAT, since a low
$Y(\gamma_*)$ leads to a high synchrotron luminosity at
high-energies. The low-energy electrons that produce early x-ray
and optical afterglow emission, however, still suffers from the
strong IC loss (i.e. $Y(\gamma_c)$ or $Y(\gamma_m)$ are typically
high) and therefore the observed early x-ray/optical afterglow
luminosity is low, compared with high-energy gamma-ray emission.

The time evolution of $Y(\gamma_*)$ also has implication for the
temporal decay slope of high-energy afterglow emission.
$Y(\gamma_*)$ increases with time and its value could be greater
than 1 at late times for some range of parameter space. This will
lead to a faster decay of the high-energy synchrotron afterglow
emission, which may explain the early fast decay of the
high-energy gamma-ray emission seen in GRB090510 and GRB090902B.

The Fermi LAT and GBM observations as well as the Swift
observations of the short burst GRB090510 are reported in Abdo et
al. (2009), Ghirlanda et al. (2009) and De Pasquale et al. (2009).
The high-energy emission above 100 MeV shows a simple power-law
decay after the peak, with a decay slope of $\alpha_{\rm
LAT}=-1.38\pm0.07$ (De Pasquale et al. 2009) or
$-1.46^{+0.06}_{-0.03}$ (Ghirlanda et al. 2009). The XRT and UVOT
starts to observe this burst from 100 s after the burst. The X-ray
spectrum is $\beta_x=0.57\pm0.08$ and the temporal decay index is
$\alpha_{x,1}=-0.74\pm0.03$ during the initial 1000 s and
subsequently steepens to $\alpha_{x,2}=-2.18\pm0.10$ with a break
at about $t_b=1.43$ ks. The spectrum and temporal decay index of
the X-ray emission before $t_b$ is well consistent with the
forward shock emission with $\nu_x<\nu_c$ and $p\simeq2.2$,
produced by a spherical blast wave expanding in a constant density
medium. The steeper decay with a slope $\alpha_{x,2}=2.18\pm0.10$
after the break is consistent with a jet break model (Kumar \&
Barniol Duran 2009). Such an interpretation predicts a decay slope
of $\alpha=\alpha_{x,1}-\Delta\alpha=-0.99$ at high-energy
frequency with $\nu_*>\nu_c$ in the standard synchrotron scenario,
since the different cooling behavior of electrons causes a
difference of $\Delta\alpha=0.25$ in the decay slope. This slope
is much shallower than the observed slope. The high-energy
gamma-ray emission observed from the long burst GRB090902B by
Fermi/LAT is reported in Abdo et al. (2009b). LAT detected
high-energy gamma-ray emission above 100 MeV on time scales much
longer than the prompt phase. The time-integrated spectrum of the
LAT detected emission after the prompt phase is consistent with
$\beta_{LAT}=-1.1\pm0.1$. Its flux declines as a simple power-law
with a decay slope of $\alpha_{LAT}={-1.5\pm0.1}$ from $t=25$ s to
1 ks. Taking $p=-2\beta_{LAT}=2.2\pm0.2$, the standard synchrotron
emission predicts a decay slope of $\alpha=-(3p-2)/4=-1.15\pm0.1$,
which is shallower than the observed decay slope, similar to the
case in GRB090510. We suggest that one possible origin for the
discrepancy in the theoretical and observed decay slopes seen in
GRB090510 and GRB090902B is due to the KN effect on high-energy
electrons{\footnote{A radiative blast wave interpretation for the
fast decay has also been proposed recently by Ghisellini et al.
(2009). However, a very high $\epsilon_e$ (i.e. $\epsilon_e\simeq
1$) in additional to the fast-cooling spectrum is needed to make
the blast wave radiative.}}, as discussed in $\S$4.

The spectrum of the afterglow emission at high-energies can be
changed as well when $Y(\gamma_*)$ increases to be large than 1 at
later times, since the electron distribution is changed in this
case. Therefore, the time-resolved spectrum at high-energies at
later times can be harder than $F_\nu\sim \nu^{-p/2}$. However,
the time-integrated spectrum of the high-energy emission will be
still $F_\nu\sim \nu^{-p/2}$ since the time-integrated fluence is
dominated by the contribution at early times (note that the flux
decays usually faster than $t^{-1}$), at which times
$Y(\gamma_*)<1$ usually. The low-significance data of the
time-resolved spectra of GRB090510 and GRB090902B prevent us from
identifying the KN effect in the spectrum (De Pasquale et al.
2009; Abdo et al. 2009b) . Obtaining high-significance
time-resolved spectra at high-energies from bright GRBs in future
would be very useful do this.

\section{Summary}
We have studied the KN effect on early high-energy  afterglow
emission of GRBs. Our findings are summarized as:

i) The IC scatterings between high-energy electrons that produce
early-time high-energy ($\ga100$ MeV) afterglow emission and
synchrotron peak photons of the afterglow  are generally in the
deep KN scattering regime. As a result, the IC loss of these
electrons is small and  the synchrotron luminosity at $h\nu\ga100$
MeV is high, which is favorable for the detection of high-energy
gamma-ray emission from the early afterglow by Fermi LAT.

ii)The high-energy gamma-ray emission at the early afterglow phase
is dominated by the synchrotron emission. The SSC emission becomes
dominated only at energies above the maximum synchrotron photon
energy of shock-accelerated electrons, but at such energies the
SSC flux is usually too weak to be detectable by Fermi LAT.

iii)The KN suppression effect of high-energy electrons weakens
with time, so that the IC loss increases with time. In the
parameter space where the Compton parameter is $Y(\gamma_*)>1$,
the increasing IC loss leads to a faster temporal decay of the
synchrotron afterglow emission at high frequency. The decay slope
could be steeper by a factor of $\Delta \alpha=0.5$ at most under
favorable conditions. This may explain the rapid decay of the
early-time high-energy emission observed in GRB090510 and
GRB090902B.

 {\acknowledgments
This work is supported by the 973 program under grants
2007CB815404 and 2009CB824800,  the NSFC under grants 10973008,
10873009, 10843007, 10503012, 10621303 and 10633040, the
Foundation for the Authors of National Excellent Doctoral
Dissertations of China, the Qing Lan Project, NCET and the NASA
grants (NNX09AT72G and NNX08AL40G).}

\end{document}